# "Playing the Robot's Advocate": Bystanders' Descriptions of a Robot's Conduct in Public Settings

Damien RUDAZ, Dept. of Economics and Social Sciences, Telecom Paris and Institut Polytechnique de Paris, France. Damien.Rudaz@telecom-paris.fr

Christian LICOPPE, Dept. of Economics and Social Sciences, Telecom Paris and Institut Polytechnique de Paris, France. Christian.Licoppe@telecom-paris.fr

## ABSTRACT

Relying on a large corpus of natural interactions between visitors and a robot in a museum setting, we study a recurrent practice through which humans "worked" to maintain the robot as a competent participant: the description by bystanders, in a way that was made accessible to the main speaker, of the social action that the robot was taken to be accomplishing. Doing so, bystanders maintained the robot's (sometimes incongruous) behaviour as relevant to the activity at hand and preserved the robot itself as a competent participant. Relying on these data, we argue that ex ante definitions of a robot as "social" (i.e., before any interaction occurred) run the risk of naturalizing as self-evident the observable result from micro-sociological processes: namely, the interactional work of co-present humans through which the robot's conduct is reconfigured as contextually relevant.

## INTRODUCTION

In the substantial body of studies focused on the co-construction of human-agent interactions, a frequent observation pertains to the amount of work required by human participants to progress the interaction with a robot (Due and Lüchow, 2022; Pelikan et al., 2022; Tuncer et al., 2023). As with self-driving cars – where the burden falls "onto other road users" (Brown et al., 2023) – or with other technologies (Greiffenhagen et al., 2023), numerous research articles have noted that human-robot interactions require "substantial adaptation from the human" (Stommel et al., 2022), during which the emergence of the robot as an agent "is contingent on participants' 'good will' to treat the robot's contributions as relevant" (Pelikan et al., 2022). Ironically, this situation was often documented in the case of caregiver robots which sometimes require more (interactional) "care" than they can provide (Chevallier, 2023; Lipp, 2022). As Alač et al. (2020) remark, "commercial social robots, designed as conversation-oriented devices, manifest their incompleteness in their need for other voices" (Alač et al., 2020).

Relying on a corpus of interactions between visitors and a robot in a museum setting, we study a specific practice through which humans "worked" to maintain the robot as a competent participant: the description by bystanders of the social action that the robot was taken to be achieving. We analyse how these descriptions of the robot's actions displayed different kinds of "footings"[1] (Goffman, 1979a; Levinson, 1988) towards the unfolding human-robot interaction, which enacted and made perceptible different stances regarding the robot as a "participant". We argue that this display of alternative footings from co-present participants was one of the many observable methods through which the robot's (sometimes incongruous) behaviour was maintained as relevant to the activity at hand (Pelikan et al., 2022) – and through which the robot itself was preserved as a competent participant.

In the end, congruently with the existing Ethnomethodological and Conversation Analytic (EMCA) literature on human-machine interactions, this micro-analytic angle leads us to a respecification of the "sociality" of the robot in a public setting. There was heavy "interactional work" (Due and Lüchow, 2022) of human participants at play behind what could otherwise be glossed as, e.g., the "wow effect" of a "social robot". That is, rather than the mechanical result of the robot's language speaking abilities, the contextual relevance of the robot's conduct was finely monitored and enforced by bystanders.

## DATA AND METHOD

### *Data collection*

Our analysis is based on a dataset recorded in July 2022 in Paris, at the Cité des Sciences et de l'Industrie, one of the biggest science museums in Europe. The first half of our corpus consists of 100 naturally occurring interactions with an autonomous Pepper robot, produced by Aldebaran[2], placed in the hall of the museum. The fragments provided in this paper took place in this natural setting. The second half is composed of 100 additional interactions which occurred in a laboratory open to the public in this museum. In both cases, participants (groups or single individuals) encountered the same Pepper robot, which was programmed to "converse" on a wide variety of topics. This dataset was collected in accordance with the General Data Protection Regulation (GDPR). Ethics approval was obtained from Paris-Saclay Université Research Ethics Board.

### *Analytic methodology*

We use an Ethnomethodological and Conversation Analytic approach, i.e., a micro-sociological level of description that studies the moment-to-moment temporal unfolding of events in an interaction. In the case of interactions involving robots, rather than dismissing single occurrences as "inherently disorderly" (Heritage, 2001), this approach attempts to grasp the typical methods and processes through which participants (human and non-human) co-produce their mutual conducts.

### *Delimitation of the phenomenon: "Playing the robot's advocate"*

In order to deepen the understanding of members' methods through which interactional "work" is accomplished in group interactions with a humanoid robot, we searched our corpus for strips of interaction in which the robot's conduct was observably maintained as relevant despite manifest interactional trouble. Among the vast array of phenomena covered by this criterion, this paper focuses on the detailed analysis of one recurring practice: the description by a bystander of the social action (interpreted as such) produced by the robot through its conduct. That is, we concentrate on multiparty interactions involving at least two humans where:

1) A co-present human – configured at this point as a bystander – provided a formulation for an action produced by the robot,
2) through which the robot's conduct was treated as *relevant and responsive to the local situation* (the sequential context, physical setting, etc.), in a way that was available to the participant to whom the robot was currently speaking.

---

[1]We use "footing" in the sense of Levinson's (1988) definition, synthetizing Goffman (1979a): the "projection of a speaker's stance towards an utterance (its truth value and emotional content)" (Levinson, 1988).
[2] https://www.aldebaran.com/

These cases display the robot's conduct being publicly turned into an action, in the sense of a "conduct under a description" (Sidnell, 2017). In other words, we did not focus on the interactional work taking place *upstream* (Chevallier, 2023) from the robot's contributions, as a preliminary scaffolding (Kamino and Sabanovic, 2023; Pelikan et al., 2022) – i.e., preconfiguring a material setting and a sequential context *where the robot has the capabilities to produce relevant contributions*. We were, instead, interested in how bystanders dealt with the robot's contributions *after they were produced*, to configure them in a way where they could be responded to by the person currently speaking with the robot – a form of a posteriori (or downstream) scaffolding of the robot's contributions.

## FRAGMENTS

*Fragment 1*

```
1. TOM       $peux tu faire une blague?
             can you tell a joke?
   rob     >>$dances-->
2.           (2.5)#
   img            #img.1
```

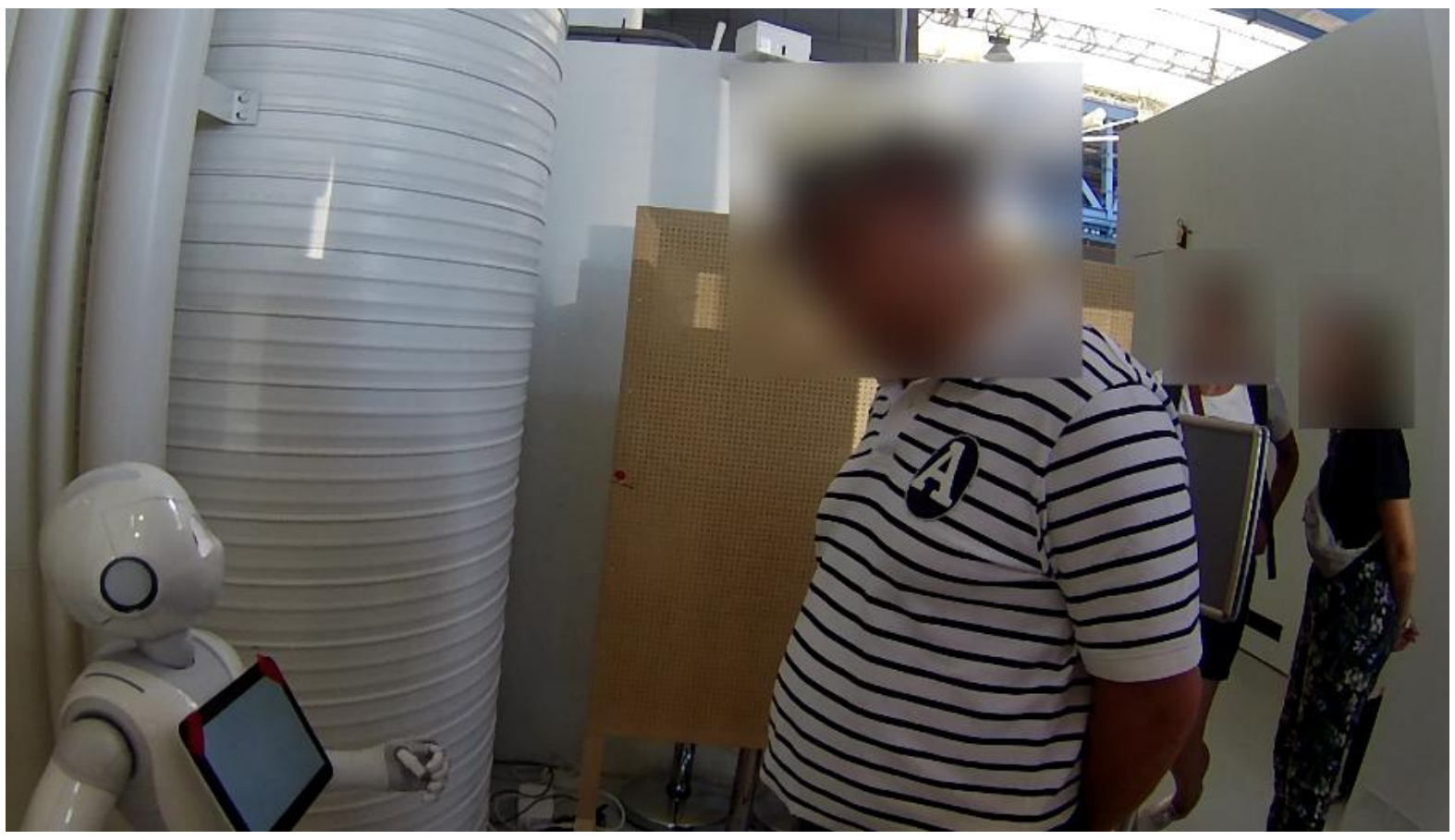

Fig.1. Img.1 ROB keeps dancing after TOM's request. SYL and CLA are positioned behind TOM as bystanders.

```
3. CLA       ah il a pas fini sa danse [hhhhh.]
             he/it didn't finish his/its dance
```

```
4.  TOM                                          [non@ mais je (...)]
                                                  no but I
    tom                                              @turns towards CLA-->
5.  SYL       tu @le pertu$rbes
              you are disturbing him/it
    tom       -->@turns towards ROB-->
    rob                -->$
6.  CLA       [h@hhh.]
    tom      -->@
7.  TOM       [hhhh]h.
8.             (0.4)
9.  ROB        si tu veux que je m'arrête dans danser (0.2) il suffit
               if you want me to stop dancing                just
10.ROB         de me dire (0.3) arrête de danser
               tell me          stop dancing
11.            (0.3)
12.TOM        °ah bah d'accord° arrête de danser
                       alright  stop dancing
13.            (1.7)
14.ROB         je ne peux pas (0.3) je ne suis pas en train de danser
               I can't              I'm not currently dancing
15.            Δ  ((choral laughter))  #Δ
    tom        Δthrows his head backwardΔ
    img                                #img.2
```

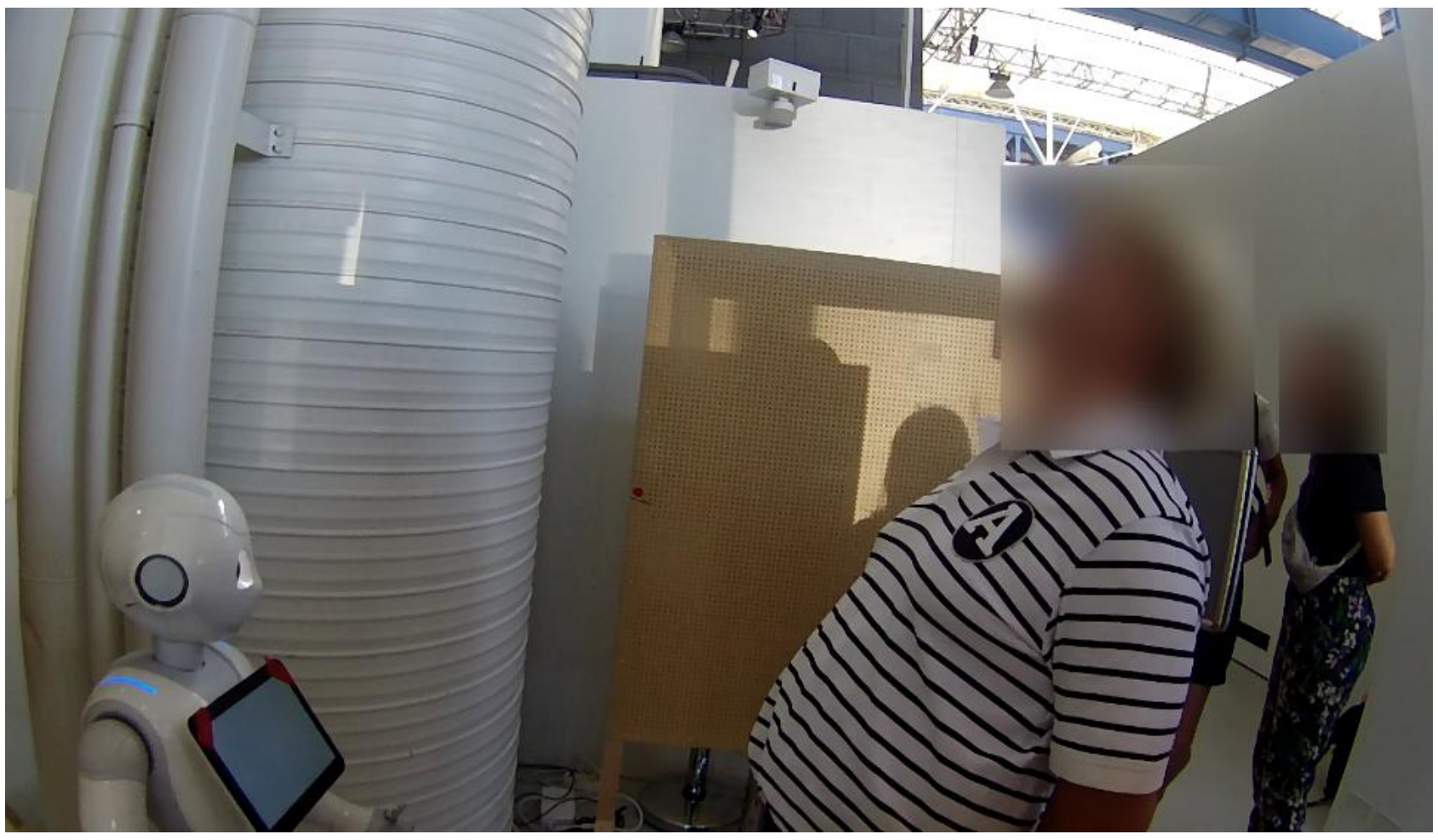

Fig.2. Img.2 TOM, SYL and CLA laugh after ROB's utterance.

```
16.CLA         y te£ troll [@(...)] [hhh£h]#  @hhh.
    ➞           he/it is trolling you
    cla            £gazes at bystander£
    tom                     @turns towards CLA@turns towards ROB
    img                                      #img.3
```

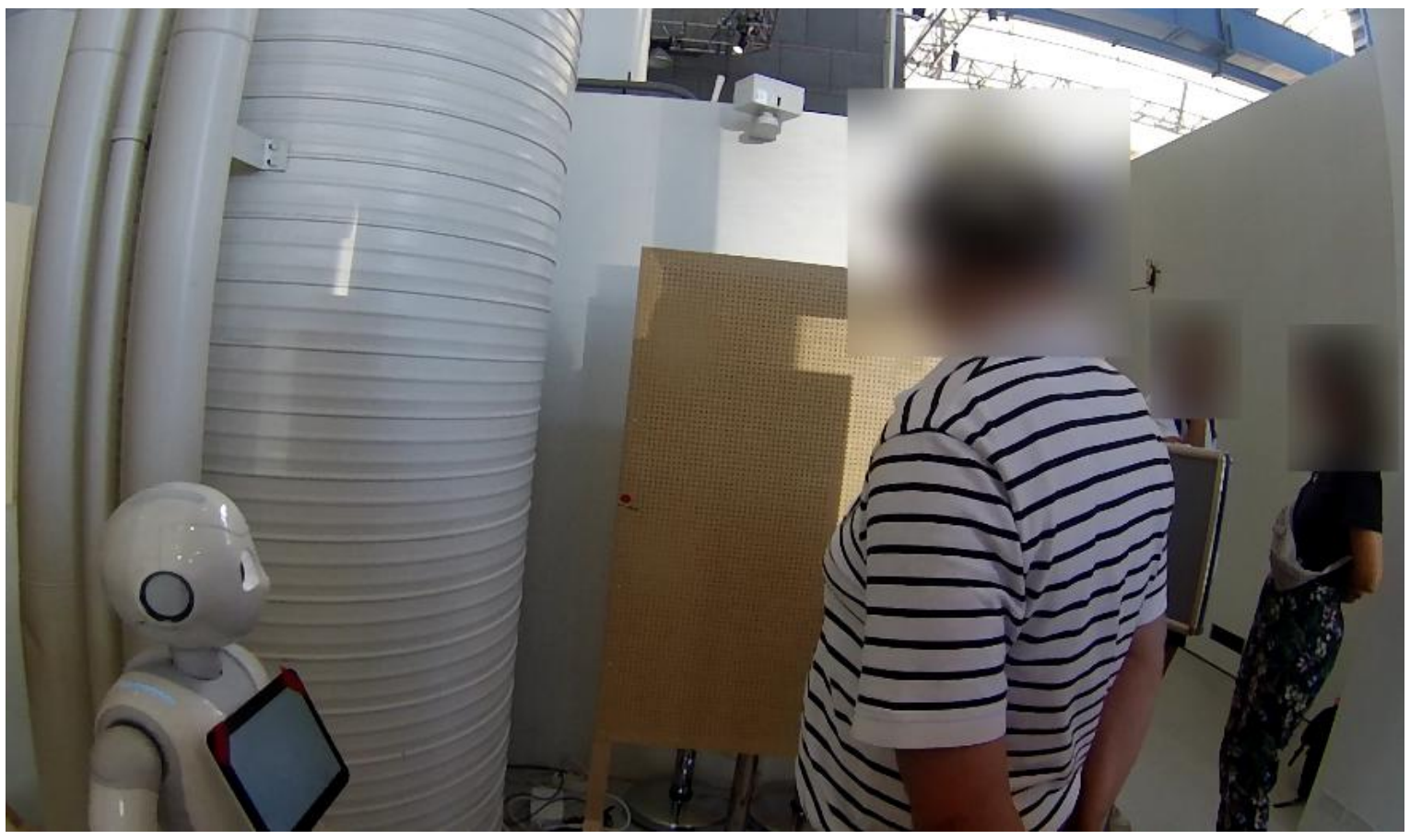

Fig.3. Img.3 After SYL and CLA describe ROB's conduct as "trolling", TOM torques towards them and produces an assessment about the situation.

```
17.ROB                          [comment?]
                                 what?
18.TOM                                   [hh (oui c'est rigolo)]
                                          yes it's funny
```

*Analysis*

While ROB stays silent after TOM's request to tell a joke (L.1), CLA and SYL, positioned behind TOM, produce explanations which "account for the robot's missing action" (Pelikan et al., 2022): "it didn't finish its dance" (L.3) and "you are disturbing it" (L.5). However, in apparent agreement with CLA and SYL's explanations about the "dancing" being what prevents it to answer, ROB states that that "if you want me to stop dancing", then "tell me stop dancing" (L.9 & 10). This utterance is produced while gazing at TOM, which strongly configures him as the addressee. TOM responds to this turn as an "informing" (Heritage, 1985) providing new and relevant information regarding the issue at hand: prefacing his turn with the french change-of-state token "ah" (Heritage, 1985), he immediately requests the robot to stop dancing (L.12). ROB's utterance is therefore reconfigured as a situational *suggestion* by TOM's response, rather than as a random bit of information.

Yet, ROB denies TOM's request to stop dancing (L.14) and produces an account for its denial: it is not currently dancing. After a choral laugh (L.15, image 2), CLA *describes the robot's turn as "trolling"*, i.e., as "baiting" or "mocking" TOM (L.16, image 3). Doing so, CLA displays a different footing towards ROB's response (L.14 & 15) and towards the immediately preceding turns: what was so far *advice-giving*, to take at face value, now gets responded to as a playful joke. This turn from CLA is the phenomenon of interest for us in this work: CLA's description of ROB's stated inability to stop dancing characterizes it as *intentionally breaching the relevancies and the expectancies* produced by its previous suggestion. "Misfitted" (Due, 2019) or "incongruous" actions have been noted to be some of the most recurring basis of "laughables" (Due, 2019) in human-robot interactions. However, here, even though participants produce a choral laughter, their conduct (playfully) reconfigures the robot's action as perfectly "fit" (rather than "misfitted") to the previous talk. Rather than an incongruous behaviour stemming from ROB's absence of understanding of the context or of its own body movements, ROB’s conduct is oriented to as the result of a detailed understanding of its co-participant’s expectancies. Bystanders build on the robot’s behaviour as a resource to construct a coherent and playful story around it.

*Fragment 2*

```
1.SAM      +@*que signifie peppeur?
              what does Pepper mean
  rob      >>+gazes at EMI>>
  sam      >>@leans forward-->
  emi      >>*holds ROB’s hand>>
2.            (3.4)%(0.2)
  rob              %displays “que signifie Pepper”-->
3.ROB         littéralement (.) ça veut dire piment au poi:vre (.)
              literally         it means chili pepper
4.ROB        £en anglais (0.5)  on £ m’a appelé comme ça parce que
              in english        I’ve been named this way because
  mir        £gazes at experimenter£-->gazes at ROB
5.ROB         je suis là pour mettre du pim%ent dans ta vie
              I’m here to add some spice in your life
  rob                                    -->%
6.MIR        £oh::/£  [hhh]
  mir     -->£gazes at experimenter-->
  emi           -->£gazes at MIR-->
7.SAM        [woa: génial: (0.7) c’est su:per]
              brilliant          this is great
8.            (0.5)£(0.3)@(.)
  emi           -->£gazes at ROB-->
  sam                  @stands straight>>
9.SAM         c’est beau hein?
              it’s beautiful huh
10.           (0.2)£#(0.7)
  sam           -->£gazes towards rest of the exhibit-->
  img               #img.4
11.EMI       £((laughs while jumping towards ROB))£
  mir     -->£gazes at ROB and EMI-->
  sam                                            -->£gazes at EMI-->
12.ROB       =pardon?
              excuse-me
13.           (0.8)£#   (0.5)   £(0.1)
  emi           -->£gazes at SAM£gazes at ROB-->
  img               #img.5
```

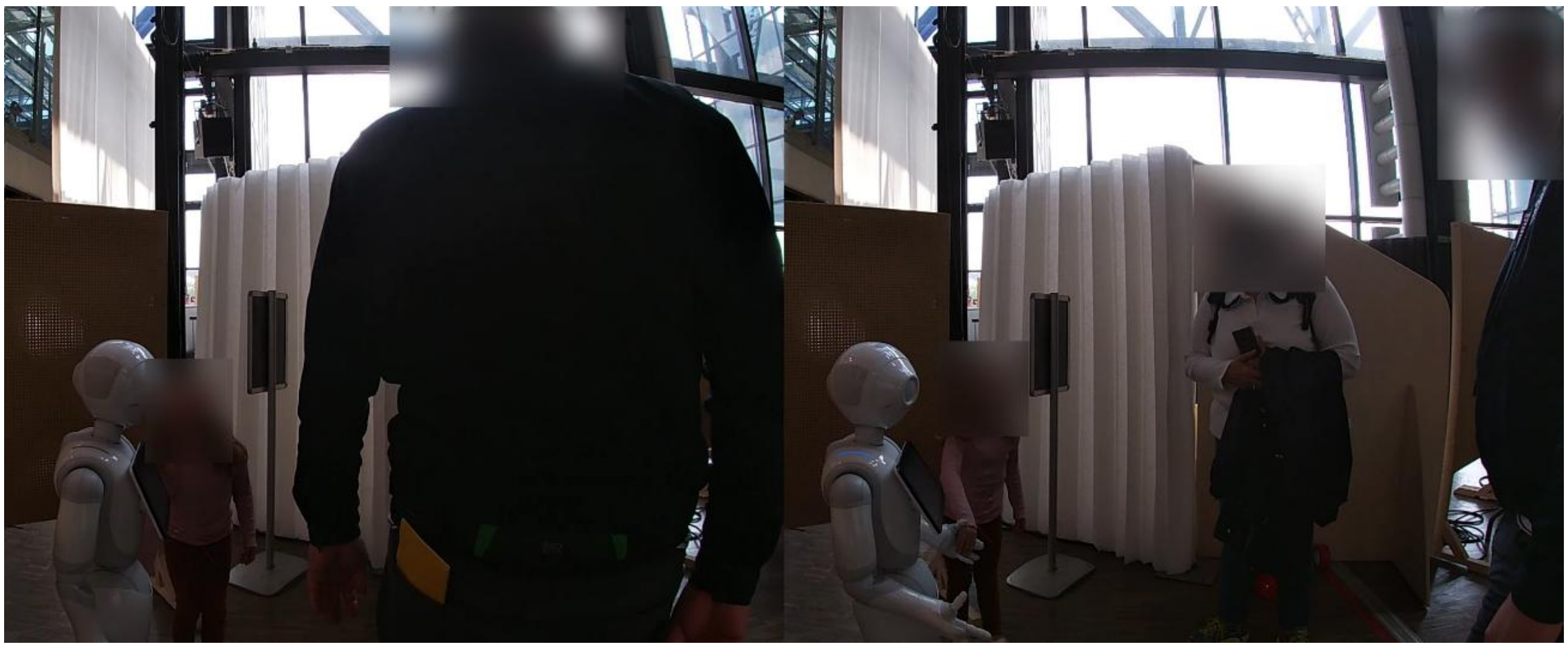

Fig.4. Img.4 & 5: EMI gazes at ROB (image 4, left), then turns towards SAM and MIR after ROB utters "pardon?" (image 5, right).

```
14.EMI      j'ai rien £[DI::::]
            I didn't say anything
15.MIR     [bah tu t'es pas]£présentée @ tu lui
            you have not introduced yourself you have not
  emi                    -->£gazes at SAM-->
  mir                                  @leans towards EMI>>
16.MIR     [as pas dit comment tu £t'a*ppelais: tu     t'appe*lais:]
            told him/it what your name was      what your name was
  mir                                 *touches EMI's shoulder*
  emi                          -->£gazes at ROB-->
17.SAM     [ouais mais faut qu'tu présentes@
            yeah but you must introduce yourself
  emi                                      @turns towards ROB-->
18.SAM     lui il s'est présenté tu t'présentes
   →       he/it introduced him/itself you introduce yourself
19.EMI     =j'm'appelle@ Emilie::
            my name is Emilie
  emi               -->@
```

*Analysis*

The beginning of this fragment features a long explanation from ROB (L.3 to L.5), responding to SAM's question about the meaning of "Pepper" (ROB's name). ROB's explanation is followed by positive assessments from MIR, through the affective particle (Hoey and Kendrick, 2017) "oh::" (L.6), and from SAM (L.7 and L.9). After ROB's response, SAM repositions himself as a bystander: he stands straight (L.8), gazes at the rest of the exhibit (L.10) and produces another assessment packaged for the audience ("it's beautiful huh?", L.9). EMI is left as the only participant forming a common interactional space (Mondada, 2009) with ROB – which holds a mutual gaze with EMI. However, following EMI's laugh (L.11, image 4), ROB produces the open-class repair initiator "excuse-me" ("pardon", L.12). In the robot's programming, uttering "excuse-me" is the consequence of hearing a strip of sound in which it couldn't identify distinct words. A silence of 1.4 seconds follows, during which EMI gazes at SAM and MIR, in a potential embodied display of trouble (Drew and Kendrick, 2018) with ROB's immediately preceding utterance (cf. image 5). EMI then looks back at ROB and screams "I said nothing" (L.14). In overlap with the end of EMI's turn, MIR produces an account of EMI's past behaviour ("you have not introduced yourself", L.15 to L.16). Also in partial overlap, SAM states that EMI should present herself (L.17 to L.18) and *describes the robot's previous turns* (L.3 to L.5) *as components of an introduction sequence* ("it introduced itself", L.18). This footing of bystanders is directly addressed to the main speaker through an embodied participation shift (L.15) which momentarily creates a common interactional space (Mondada, 2009) between all three human participants. Finally, EMI aligns with SAM and MIR's stance by identifying herself towards ROB (L.19).

Both turns (L.15 to L.18) from SAM and MIR therefore orient to EMI's conduct as normatively accountable for not having produced a sequentially expected next move: she didn't "tell it what her name was" (L.16)**.** This retrospectively configures ROB's explanation about the origin of its name (L.3 to L.5) as launching a mutual identification sequence (Schegloff, 1986) and, thereby, as producing specific expectancies for ROB's interlocutor. Note that participants' accounts of the situation can themselves be analysed as progressing an overarching educational activity directed towards EMI: akin to a step-by-step tutorial, SAM specifies the normative expectation made relevant by the robot's original turn and its adequate response – "it introduced itself you introduce yourself" (L.18). Even though this is not made hearably relevant by bystanders, their account is possibly reinforced by ROB's gaze orientation: it has been gazing at EMI since the beginning of the interaction, providing an additional cue that its "self-identification" (L.3 to L.5) was addressed to EMI, and not merely a factual response to SAM's original question (L.1). Hence, ROB's conduct is treated by both bystanders as resulting from a detailed understanding of the sequential expectancies produced by its previous conduct. ROB's open-class repair initiator "excuse-me" (L.12) is not responded to as indicating a problem of hearing but as making noticeable EMI's non-response to ROB's immediately preceding talk. In other terms, we argue that this fragment should not be analysed as a simple case of "alignment" of humans with the initiation of an identification sequence by the robot. Instead, the whole identification sequence *is retroactively produced by bystanders to fit the emerging conduct of the robot.*

## DISCUSSION

*The mutual shaping of the robot's conduct*

In the setting of a museum exhibition, bystanders regularly described what the robot was doing or responding to, and did so in a way that was publicly accessible to the main speaker. These contributions displayed specific footings towards the robot's conduct (as producing contextually relevant jokes, as manifesting normative expectations as part of an identification sequence, etc.) that made these conducts publicly available as relevant to the task at hand – and that shaped the main speaker's response to these conducts. As Alač et al. (2020) remark, following Goffman's (1979a) participation framework, "talk as action does not concern speakers alone, but is a product of interaction, where coparticipants (who, as bystanders and non-ratified participants, may even not be speaking) play an active role" (Alač et al., 2020).

Significantly, this work to make technology "work" (Greiffenhagen et al., 2023) was not only produced by the robot's co-present and officially appointed "caretakers" (experimenters, demonstrators, care workers; Chevallier, 2023). Accounting for the robot's conduct (i.e., the "work" of making the robot's conduct both *recognizabl*e and *relevant*; Robinson, 2016) was partially supported by human co-participants themselves – even by those who were not, in a different vocabulary, the "end-users" of the dyadic interactions the robot was designed for. Although bystanders did not "back up" (Pitsch, 2020) the main speaker's responses to the robot, their impact was akin to that of a "participation facilitator" (Pitsch, 2020) who, simultaneously, preserved the robot as a competent participant. However, that bystanders participated in framing the robot's conduct as relevant (i.e., the observable impact of their contributions on the course of the talk) does not entail that they were doing "not-doing repair" (Pilnick et al., 2021) in the sense of *intentionally* avoiding repair (Stommel et al., 2022) when the relevance of the robot's contribution was not "immediately discernible" (Pilnick et al., 2021) even to them as bystanders – and not just to the main speaker.

*The robot as a tool to advance local human activities*

Crucially, descriptions of the robot's conduct were not a "mere summary" (Antaki, 2008) stemming from a neutral observer removed from the urgency and the relevancies of the local interaction but, rather, they were a contribution to an ongoing and collective activity – in which the bystander producing the description was caught up. For example, participants' description of the robot's behaviour as a self-introduction (fragment 2) takes place as part of the observable educational endeavour of "teaching manners" or "values" (Tannen, 2004), i.e., accustoming a child to identify herself. Hence, in our corpus, the "sociality" of the robot was, often, an expedient tool: through the descriptions that participants produced of its conduct (among other practices), the robot was produced as "social" by human participants as a resource, or a "toy" (Goffman, 1979b; Roberts, 2004), to accomplish various local "interactive goals" (Tannen, 2004).

*Side-sequences: a typical practice in public human-robot interactions*

In an apparent paradox, bystanders' contributions which maintained the robot as a competent participant took place during side-sequences (Jefferson, 1972) from which the robot was excluded as a participant. Indeed, during these side-sequences – congruently with those identified by Krummheuer (2015) in multiparty human-agent interactions taking place in a shopping centre – the main speaker (i.e., the one who initially took the "stage" in front of an audience; Krummheuer, 2015) torqued or fully turned towards bystanders. Doing so, this main speaker established a momentary focused interaction with bystanders, during which the robot was often discussed using the third person "it" or "he" in our corpus. Hence, through these side-sequences, the robot was "not given a chance to face a problematic issue in the communication" (Krummheuer, 2015). Similarly, studying public interactions with a Pepper robot, Due (2019) describes the formation of spatial configurations which "exclude Pepper from the sense-making (but not the spatial) framework". The presence and the format of these participation shifts therefore appear not to be specific to our corpus and, instead, to be typical of the activity of interacting with a robot as a group in a public space. Because they excluded the robot as a participant, these side-sequences constitute a specific case of the fast-paced changes in status often undergone by non-human agents (Pelikan et al., 2022), which can be "enacted, in one breath, as an agent and a thing" (Alač, 2016).

## THE INNER WORKINGS OF THE ROBOT'S "SOCIALITY"

Attending to the moment-to-moment production of the robot's status as a practical accomplishment (Rollet et al., 2017) – as done by the micro-analytic studies cited over the course of this paper – leads to a respecification of the lay use of the concept of "social robots". The mostly positive encounters that we studied did not mechanically arise because humans faced a humanoid and intended-to-be-social robot for the first time which, e.g., produced a "wow effect"[3]. Such a summary would suppress the collaborative work produced by co-present humans to enforce the robot's conduct as relevant. Instead, the robot was observably maintained as a "social robot" *in spite* of the recurring hard-to-interpret talk (Pilnick et al.,

[3] Applied to robots, the "wow effect" has been used as a lay concept to describe the observable tendency of humans to have positive first encounters with a robot (Etemad-Sajadi et al., 2022).

2021) it produced for the main speaker.

Ex ante definitions of the robot – or any technology – as “social” (i.e., before any interaction occurred) therefore run the risk of naturalizing, as stable (Pollner, 1974) and self-evident, the observable result from micro-sociological, and always fallible, processes. In the case of the humanoid robot used in the fragments above, a definition of the robot’s “sociality” as the unmediated consequence of its technical abilities (being able to speak, to move its arms, to detect and to respond to humans, etc.) – rather than as a locally emerging property of the interaction (Licoppe and Rollet, 2020) – obscures the micro-interactional work that bystanders achieved to enforce the robot’s behaviors as “social actions”: that is, as actions responsive to the situated interaction and as “making relevant a set of potential next actions” (Tuncer et al., 2022). As we attempted to show, used in their lay sense, notions of “social” or “conversational” robots “are merely ‘glosses’ for the processes that constitute them” (Hoeppe, 2023).